\documentclass[12pt]{article}
\pdfoutput=1

\usepackage[pdftex]{graphicx}
\usepackage{amssymb}
\usepackage{amsmath}
\usepackage{cite}
\usepackage{multirow}
\usepackage{gensymb}
\usepackage{longtable}
\usepackage{booktabs}
\usepackage{array}
\usepackage{anyfontsize} 
\usepackage{ulem}
\usepackage{wrapfig}
\usepackage{xcolor}
\usepackage{cancel}
\usepackage{arydshln}

\definecolor{tit}{rgb}{0.1,0.2,0.4}

\newcommand{\bea}{\begin{eqnarray}}
\newcommand{\eea}{\end{eqnarray}}
\newcommand{\be}{\begin{equation}}
\newcommand{\ee}{\end{equation}}
\newcommand{\non}{\nonumber\\ }

\definecolor{DRed}{rgb}{0.8,0,0.1}
\definecolor{DBlue}{rgb}{0,0,0.8}

\newcommand{\blue}{\color{blue}}

\newcommand{\gray}{\color{gray}}

\begin{document}

\vspace{-2mm}
\begin{center}
\fontsize{20}{18}
\selectfont
\bf
Dipion light-cone distribution amplitudes and $B \to \pi\pi$ form factors
\end{center}

\vspace{-2mm}
\begin{center}
\large
{\sf Shan Cheng }  \\[3mm]
{\sf \small
School of Physics and Electronics, Hunan University, \\[1mm]
410082 Changsha, People's Republic of China.} \\

\end{center}


\date{\today}

\begin{abstract}
We suggest to update the expansion coefficients of 2$\pi$DAs with the distribution amplitudes of light mesons evaluated from lattice QCD, 
with which we revisit $\overline{B}^0 \to \pi^+\pi^0$ transition form factors from light-cone sum rules approach and 
extend the predictions from the threshold of dipion invariant mass to high energies with including the resonance intervals.  
We also derive $B^- \to \pi^0\pi^0$ transition form factors with the isoscalar dipion final state, 
serving as the supplement to the isovector ones to complete the set of light-cone sum rules prediction of $B \to \pi\pi$ form factors. 
Our numerics shows that the lowest resonance gives the dominant contribution to $P-$wave form factors, 
while the resonance contribution in $S-$wave is not so salient.

\end{abstract}

\allowdisplaybreaks

\newpage

\tableofcontents

\bigskip

\section{Introduction}\label{sec:introduction}

With a large number of angular observables that is sensitive to the spin structure of the underlying short distance operators, 
the $B \to \pi\pi \ell \overline{\nu}_\ell$ decays had been suggested to probe the $V-A$ nature of weak interaction\cite{FallerDWA}, 
whose measurement would provides prolific information to test QCD theoretical approaches and to search physics beyond the SM. 
At the quark level, it is generated by the semileptonic $b \to u\ell \overline{\nu}_\ell$ transition, 
which offers another independent channel to determine the Cabiboo-Kobayashi-Maskawa matrix element $V_{ub}$ 
if we are able to calculate the $B \to \pi\pi$ form factors with an adequate accuracy.  

There are some efforts on this topic recently. 
At large dipion invariant mass, the QCD factorization is available\cite{BoerIEZ,FeldmannKQR},  
when the invariant mass is small and the hadronic recoil is low,  
the chiral effective theory of heavy meson is proposed to combine with the dispersion theory\cite{KangJAA}. 
The light-cone sum rules (LCSRs) approach is operative in the regions of large hadronic recoil at low invariant mass, 
i.e., the $S-$wave generated $B \to \pi\pi, K\pi$ form factors has been calculated\cite{MeissnerHYA,MeissnerPBA} with 
the combination of the perturbative theory based on the operator production expansion and the low-energy effective theory inspired by the chiral symmetry, 
corresponding to the effect from the LCDAs and from the scalar form factors, respectively, 
and recently the $\overline{B}^0 \to \pi^+\pi^0$ form factors has been derived from LCSRs approach 
with the generalized 2$\pi$DAs\cite{HambrockAOR,ChengSFK,ChengSMJ}.  
In principle, the low invariant mass in LCSRs approach to deal with $B\to\pi\pi \ell \overline{\nu}_\ell$ decays 
goes from the threshold $4m_\pi^2$ to the resonance intervals, 
the previous works follow closely around the threshold while the running on the invariant mass is less attended, 
but actually this part is indispensable to show the resonance contribution. 
The LCSRs prediction is always influenced by the high power terms, 
one of which is the non-asymptotic QCD corrections of hadron distribution amplitudes (DAs), 
in literatures, most of the calculations of $B\to\pi\pi$ form factors use the expansion coefficients of 2$\pi$DAs 
obtained 20 years ago from instanton model\cite{PolyakovZE}.

In this paper we suggest to update the non-asymptotic coefficients of leading twist 2$\pi$DAs with the lattice QCD (LQCD) knowledge of light mesons,  
which is already quite accurate with the development of the discrete computing technique. 
We prolong the previous calculation of $\overline{B}^0 \to \pi^+\pi^0$ form factors from the threshold of dipion invariant mass 
to the rather broad range with the energy-dependent 2$\pi$DAs, 
we also derive the new LCSRs prediction for the $B^- \to \pi^0\pi^0$ form factors with the isoscalar dipion final state, 
which is an essential physical quantity in the angular observables of the semileptonic decay.
All calculations in this paper are at leading twist and the contributions from high twist 2$\pi$DAs are postponed for the future work.

The paper is organized as follows.
In section.\ref{sec:2piDAs}, 
we discuss briefly the properties of 2$\pi$DAs and update the coefficients of isovector 2$\pi$DAs with the LQCD result of light mesons. 
Section.\ref{sec:B2pipi_ff} is the main part of this paper, 
where we present the LCSRs' calculation of $B \to \pi\pi$ form factors with both the isovector and isoscalar dipion final states.
Section.\ref{sec:numerics} is the numerical result.
Our conclusions are presented in Sec.\ref{sec:conclusion}.

\section{Dipion light-cone distribution amplitudes}\label{sec:2piDAs}

The 2$\pi$DAs are the most general object to describe the dipion mass spectrum in hard production processes, 
whose asymptotic formula indicates the information about the deviation of the unstable meson DAs($\rho, f_0, a_0$ etc.), 
and in further to improve the theoretical accuracy of the nonperturbative information of meson, like the decay width. 
From the other hand, the crossing symmetry implies a relation between 2$\pi$DAs and the skwed parton distributions (SPDs) in the pion, 
which provide another constraint to determine the SPDs.
In this section we brief discuss the properties of 2$\pi$DAs, and explain some updates of the nonperturbative inputs. 
Our precise is still at leading twist. 

\subsection{General review of 2$\pi$DAs}\label{subsec:2piDAs}

The concept of wave function of a single meson has been generalized to a multi-hadron system\cite{BaierVW}, 
and the perturbative behavior of meson pairs is calculated at large invariant mass\cite{Grozin:1983tt,Grozin:1986at},  
while the factorized form of exclusive electroproduction process\cite{Diehl-1998,Diehl-2000} 
involves the 2$\pi$DAs at small invariant mass\cite{Diehl-2002,Diehl-2003}. 
In this work we quote the chirally even and odd two quark 2$\pi$DAs defined in Refs.\cite{PolyakovZE,LehmannDronkeAQ} as,
\bea
&&\langle \pi^a(k_1)\pi^b(k_2) \vert \overline{q}_f(xn) \gamma_\mu  \tau q_{f'}(0) \vert 0 \rangle = 
\kappa_{ab} \, k_\mu  \int dx \, e^{i zx(k\cdot n)} \, \Phi^{ab,ff'}_{\parallel}(z,\zeta,k^2) \,, 
\label{eq:Phi-para}\\
&&\langle \pi^a(k_1)\pi^b(k_2) \vert \overline{q}_f(xn) \sigma_{\mu\nu} \tau q_{f'}(0) \vert 0 \rangle= 
\kappa_{ab} \, \frac{2i}{f_{2\pi}^{\perp}} \frac{k_{1\mu}k_{2\nu} - k_{2\mu}k_{1\nu}}{2\zeta - 1} \int dx \, e^{i zx(k\cdot n)} \, \Phi^{ab,ff'}_{\perp}(z,\zeta,k^2)  \,, 
\label{eq:Phi-perp}
\eea
respectively, where the index $f,f'$ respects the (anti-)quark flavor, 
$a,b$ indicates the electro charge of each pion, 
the coefficient $\kappa_{+-/00}=1$ and $\kappa_{+0}=\sqrt{2}$, 
$k=k_1+k_2$ is the invariant mass of dipion state, $n^2=0$, 
$\tau = 1/2, \tau^3/2$ corresponds to the isoscalar and isovector 2$\pi$DAs, respectively, 
the chirally odd constant $f_{2\pi}^\perp$ is defined by the local matrix element, 
\be
\lim_{k^2 \to 0} \langle \pi^a(k_1) \pi^b(k_2) \vert \overline{q}(0) \sigma_{\mu\nu} \frac{\tau^3}{2} q(0) \vert 0 \rangle = 
\frac{2i \, \epsilon^{ab3}}{f_{2\pi}^\perp} (k_{1\mu}k_{2\nu} - k_{2\mu}k_{1\nu}) \, .
\ee
The generalized 2$\pi$DAs depend on three independent kinematic variables, 
the momentum fraction $z$ carried by anti-quark with respecting to the total momentum of dipion state, 
the longitudinal momentum fraction carried by one of the pions $\zeta = k_1^+/k^+$, 
and the invariant mass squared $k^2$. 
The normalization conditions of the distribution amplitudes read as:
\bea
&&\int_0^1 \, dz \, \Phi_{\parallel, \perp}^{I=1}(z,\zeta,k^2) = (2 \zeta -1) F_{\pi, t}(k^2) \,, \\
\label{eq:norm-vector}
&&\int_0^1 \, dz \, (2z-1) \Phi_{\parallel}^{I=0}(z,\zeta,k^2) = -2 M_2^{(\pi)} \zeta (1 - \zeta) F^{\textrm{EMT}}_\pi(k^2) \,,
\label{eq:norm-scalar}
\eea
where $F_\pi(k^2)$ is the timelike pion form factor, $F_t(k^2)$ is the tensor pion form factor, normalized by $F_\pi(0) =1$ and $F_t(0) = 1$,  
$M_2^{(\pi)}$ is the momentum fraction carried by quarks in the pion associated to the usual quark distribution \cite{JiPC}, 
$F^{\textrm{EMT}}(k^2)$ is the form factor of the quark part of the energy momentum tensor with the normalization $F^{\textrm{EMT}}_\pi(0)=1$ \cite{PolyakovEXB}.
The 2$\pi$DAs 
can be decomposed, respecting to flavor/isospin,  
as\footnote{Here we do not take in to account the 
isoscalar 2$\pi$DAs generated from two gluon configurations, 
whose contributions to a hard process are actually at next-to-leading-order 
and should be considered together with the one-loop correction to the quark 2$\pi$DAs, 
which exceeds the scope of this work.}:
\bea
&&\Phi^{ff'}_{\pi^+\pi^-}(z,\zeta,k^2) = \delta^{ff'} \, \Phi^{I=0}(z,\zeta,k^2)  + \tau_3^{ff'} \, \Phi^{I=1}(z,\zeta,k^2) \,, \non
\label{eq:Phipi+pi-}
&&\Phi^{ff'}_{\pi^+\pi^0}(z,\zeta,k^2) = \tau_3^{ff'} \, \Phi^{I=1}(z,\zeta,k^2) \,, \non
\label{eq:Phipi+pi0}
&&\Phi^{ff'}_{\pi^0\pi^0}(z,\zeta,k^2) = \delta^{ff'} \, \Phi^{I=0}(z,\zeta,k^2) \,.
\label{eq:Phipi0pi0}
\eea

As is well known, 2$\pi$DAs can be double decomposed in terms of Gegenbauer polynomials $C_n^{3/2}(2z-1)$ (eigenfunction of the evolution equation) 
and the Legendre polynomials $C_\ell^{1/2}(2\zeta-1)$ (partial wave expansion), 
\bea
&&\Phi^{I=1}(z,\zeta,k^2, \mu) = 6z(1-z) \sum_{n=0,\textrm{even}}^{\infty} \sum_{l=1,\textrm{odd}}^{n+1} \, 
B_{n\ell}^{I=1}(k^2, \mu) C_n^{3/2}(2z-1)C_\ell^{1/2}(2\zeta-1) \,, \\
\label{eq:Phi-vector}
&&\Phi^{I=0}(z,\zeta,k^2, \mu) = 6z(1-z) \sum_{n=1,\textrm{odd}}^{\infty} \sum_{l=0,\textrm{even}}^{n+1} \, 
B_{n\ell}^{I=0}(k^2, \mu) C_n^{3/2}(2z-1)C_\ell^{1/2}(2\zeta-1) \,.
\label{eq:Phi-scalar}
\eea
The coefficients $B_{n\ell}(k^2,\mu)$ have the similar scale dependence as the gegenbauer moments of pion and rho mesons \cite{ChernyakAS,LepageZB},
\bea
B_{n\ell}(k^2, \mu) = B_{n\ell}(k^2, \mu_0) \left[\frac{\alpha_s(\mu)}{\alpha_s(\mu_0)}\right]^{(\gamma_n^{(0)}-\gamma_0^{(0)})/(2\beta_0)} \,,
\label{eq:Bnl-scale}
\eea
where $\beta_0 = 11 - 2N_f/3$ and the one-loop anomalous dimension are \cite{GrossCS}
\bea
&&\gamma_n^{\parallel,(0)} = 8 C_F \left( \sum_{k=1}^{n+1} \frac{1}{k} - \frac{3}{4} - \frac{1}{2(n+1)(n+2)} \right) \,, \non
&&\gamma_n^{\perp,(0)} = 8 C_F \left( \sum_{k=1}^{n+1} \frac{1}{k} - \frac{3}{4} \right) \,.
\label{eq:gamma-n}
\eea
We note that for isovector (isoscalar) 2$\pi$DAs, the gegenbauer index $n$ goes over even (odd) 
and the partial-wave index $l$ goes over odd (even) numbers, which is guaranteed by the C-parity. 
Several comments are in time for the expansion coefficients $B_{n\ell}(k^2)$:
\begin{description}
\item[$\ddag$] 
When the four momentum of one of the pions goes to zero, soft pion theorem relates the chirally even coefficients  
and the  gegenbauer moments of pion meson,
\be
\sum_{\ell=1}^{n+1} B_{n\ell}^{\parallel,I=1} = a_n^\pi \,, \,\,\,\,\,\, \sum_{\ell=0}^{n+1} B_{n\ell}^{\parallel, I=0} = 0 \,.
\label{eq:api-Bnl}
\ee
\item[$\ddag$]
The 2$\pi$DAs are also related to the skewed parton distributions (SPDs) in the pion by the crossing, 
which support us to express the moments of SPDs in terms of $B_{nl}(k^2)$ in the forward limit as
\bea
&&M_N^\pi = \frac{3}{2} \frac{N+1}{2N+1}B_{N-1,N}^{I=1}(0) \,,\,\,\, \textrm{for odd N} \,, \\
\label{eq:cr-vector}
&&M_N^\pi = 3 \frac{N+1}{2N+1}B_{N-1,N}^{I=0}(0) \,,\,\,\, \textrm{for even N} \,,
\label{eq:cr-scalar}
\eea
\item[$\ddag$]
The Watson theorem of pion-pion scattering amplitudes implies a intuitive way to express the imaginary part of 2$\pi$DAs, 
which subsequently deduces the Omn{\'e}s solution of $N-$subtracted dispersion relation for the coefficients, 
\bea
B_{n\ell}^{I}(k^2) = B_{n\ell}^{I}(0) \, \textrm{Exp}\left[ \sum_{m=1}^{N-1} \frac{k^{2m}}{m!} \frac{d^m}{dk^{2m}} \ln B_{n\ell}^I(0) +
\frac{k^{2N}}{\pi} \, \int_{4m_\pi^2}^\infty ds \, \frac{\delta_\ell^I(s)}{s^N (s-k^2-i0)} \right] \,.
\label{eq:Bni-dr}
\eea
With two subtraction, this expression gives an excellent description of the experimental data of pion form factor 
not only below the inelastic threshold $k^2<16m_\pi^2$, but also in the resonance region up to $k^2 \sim 2.5 \, \textrm{GeV}^2$. 
In this way, 2$\pi$DAs in a wide range energies is given by the $\pi\pi$ phase shift $\delta_\ell^I$ and a few subtraction constants. 
\item[$\ddag$]
Taking the vanishing width limit in the vicinity of resonance, 2$\pi$DAs reduce to the distribution amplitudes of resonance ($\rho$), 
which implies another relation between the gegenbauer moments of rho meson and the coefficients $B_{n\ell}$,
\bea
&&a_n^\rho = B_{n1}(0) \, \textrm{Exp} \left[ \sum_{m=1}^{N-1} c_m^{n1} m_\rho^{2m} \right] \,,\,\,\,\,\,\,
c_m^{(n1)} = \frac{1}{m!} \frac{d^m}{dk^{2m}} \left[ \ln B_{n1}(0) - \ln B_{01}(0) \right] \,.
\label{eq:arho-Bnl}
\eea
The decay constants of resonance are related to the imaginary part of $B_{nl}(m_\rho^2)$ as 
\be
f_\rho^\parallel = \frac{\sqrt{2} \, \Gamma_\rho \, \textrm{Im} B_{01}^\parallel(m_\rho^2)}{g_{\rho\pi\pi}} \,, \,\,\,\,\,\,
f_\rho^\perp =  \frac{\sqrt{2} \, \Gamma_\rho \, m_\rho \, \textrm{Im} B_{01}^\perp(m_\rho^2)}{g_{\rho\pi\pi} \, f_{2\pi}^{\perp}} \,,
\label{eq:frho-Bnl}
\ee
with the strong coupling defining in $\langle \pi(k_1) \pi(k_2) \vert \rho \rangle = g_{\rho\pi\pi} (k_1-k_2)^\alpha \epsilon_\alpha $.
\end{description}

\subsection{Some remarks on the coefficients $B_{nl}(k^2)$ }\label{subsec:2piDAs_updata}

The first several subtraction constants of $B_{nl}(k^2)$ are calculated in the effective low-energy theory 
based on instanton vacuum at the normalization scale $\mu \sim 1/\overline{\rho} \approx 600 \, \textrm{MeV}$ \cite{PolyakovZE,PolyakovTD}, 
with $\overline{\rho}$ being the average instanton size. 
In Tab.\ref{tab1-Bnl}, we present their result for both the isovector and isoscalar 2$\pi$DAs, referring to $\mu = 1 \, \textrm{GeV}$.  
The values supplemented and stressed in blue are the updated result obtained by using the constraints addressed in the previous subsection, 
for the input values of $a_2^\pi \,, a_2^\rho \,, f_{\rho}^\parallel \,, f_\rho^\perp$, we adopt the lattice evaluation and the experiment measurement. 

\begin{table}[tb]
\caption{The subtraction constants of $B_{n\ell}(s)$ in Eq.(\ref{eq:Bni-dr}).}
\begin{center}
\begin{tabular}{c|c c c|c c c}
\toprule
(nl) & $B_{n\ell}^{\parallel}(0)$ & $c_1^{\parallel,(nl)}$ & $\frac{d}{dk^2} \ln B_{n\ell}^{\parallel}(0)$ &
$B_{n\ell}^{\perp}(0)$ & $c_1^{\perp,(nl)}$ & $\frac{d}{dk^2} \ln B_{n\ell}^{\perp}(0)$  \\
\toprule
(01) & 1       & 0 & 1.46 {\blue $\to$ 1.80}  & 1      & 0       & 0.68 {\blue $\to$ 0.60} \\
(21) & -0.113 {\blue $\to$ 0.218}    & -0.340 & 0.481  & 0.113 {\blue $\to$ 0.185} & -0.538 & -0.153 \\
(23) & 0.147 {\blue $\to$ -0.038}  & 0       & 0.368  & 0.113 {\blue $\to$ 0.185} & 0       & 0.153 \\
\hline\hline
(10) & -0.556    & -   & 0.413  & -    & -   & -                \\
(12) & 0.556     & -   & 0.413  & -    & -   & -                 \\
\hline
\end{tabular}
\end{center}
\label{tab1-Bnl}
\end{table}

\begin{table}[tb]
\caption{The Gegenbauer moment $a_2^\pi$ at scale $\mu = 2 \, \textrm{GeV}$.}
\begin{center}
\begin{tabular}{c|c|c }
\toprule
\textrm{Method}  & $a_2^\pi(2 \, \textrm{GeV})$ & \textrm{Refs} \\ 
\hline
LO QCDSR, CZ model  & 0.39 & \cite{ChernyakEJ,ChernyakZZ} \\
\hline
QCDSR          & $0.18^{+0.15}_{-0.26}$ & \cite{KhodjamirianGA} \\
QCDSR          & $0.19 \pm 0.06$            & \cite{BallWN} \\
QCDSR,NLC  & $0.13 \pm 0.04$            & \cite{MikhailovPT,BakulevPF} \\
\hline
$F_{\pi\gamma\gamma^\ast}$, LCSRs                    & $0.12 \pm 0.04 (2.4 \, \textrm{GeV})$ & \cite{SchmeddingAP} \\
$F_{\pi\gamma\gamma^\ast}$, LCSRs                    & $0.21(2.4 \, \textrm{GeV})$                 & \cite{BakulevUC} \\
$F_{\pi\gamma\gamma^\ast}$, LCSRs, R               & $0.19                                          $                 & \cite{AgaevRC} \\
$F_{\pi\gamma\gamma^\ast}$, LCSRs, R               & $0.31                                          $                 & \cite{BakulevCP} \\
$F_{\pi\gamma\gamma^\ast}$, LCSRs, NLO          & $0.096                                         $                 & \cite{AgaevAQ} \\
$F_{\pi\gamma\gamma^\ast}$, LCSRs, NLO          & $0.068                                         $                 & \cite{AgaevTM} \\
\hline
$F_{\pi}^{em}$, LCSRs         & $0.17 \pm 0.10 \pm 0.05$ & \cite{BraunUJ} \\
$F_{\pi}^{em}$, LCSRs, R    & $0.14 \pm 0.02$                 & \cite{AgaevGU} \\
\hline
$F_{B\to \pi}$, LCSRs           & $0.13 \pm 0.13$                 & \cite{BallTB1} \\
$F_{B\to \pi}$, LCSRs           & $0.11$                                & \cite{DuplancicIX,KhodjamirianUB} \\  
\hline
LQCD, TWST, $N_f=2$, CW         & $0.201 \pm 0.114$ &  \cite{BraunDG} \\
LQCD, TWST, $N_f=2+1$, DWF   & $0.233 \pm 0.088$ &  \cite{ArthurXF} \\
{\blue LQCD, MST, $N_f=2$}        &  {\blue $0.136 \pm 0.03$} &   {\blue \cite{BraunAXA}} \\
{\gray LQCD, MST, $N_f=2+1$, CW}       & {\gray $0.0762 \pm 0.0127$} &   {\gray \cite{BaliUDE}} \\
\hline
\end{tabular}
\end{center}
\label{tab2-a2pi}
\end{table}

Implantation of the soft pion theory and the resonance approximation in Eqs. (\ref{eq:api-Bnl},\ref{eq:arho-Bnl}) 
requires the precise input for $a_n^\pi$ and $a_n^\rho$, 
we present their values for $n=2$ obtained in different methods in Tab.\ref{tab2-a2pi} and Tab.\ref{tab3-a2rho}, 
respectively\footnote{The moments $a_n^{\pi,\rho}$ vanish with odd $n$ in the isospin symmetry limit.} .
Both the QCD sum rules and the lattice QCD (LQCD) methods were used for the direct calculation of the second moments, 
the constraints for $a_2^\pi$ come from fitting the experiment data with the LCSRs calculations for 
pion transition, electromagnetic from factors and also $B \to \pi$ form factors, 
while the constraints for $a_2^\rho$ are much less due to the width effects. 
Recently, they are evaluated in LQCD with using two flavours of dynamical fermions on lattices of different volumes ($N_f=2$) 
and pulling the pion masses down to almost physical value ($m_\pi=150 \, \textrm{MeV}$) 
\cite{BraunAXA,BraunWNX}\footnote{The momentum smearing technique (MST) is recently proposed for hadronic interpolators
to improve the lattice calculations of matrix elements of local operators involving covariant derivatives \cite{BaliUDE}, 
but the quoted low value of $a_2^\pi$ there should not be used for phenomenology 
because it is purely a methodical work and the evaluation is not completely.}.  
In the following we pick  
$a_2^\pi(1 \, \textrm{GeV}) = 0.180, \, a_2^{\rho,\parallel}(1 \, \textrm{GeV}) = 0.177, \, a_2^{\rho,\perp}(1 \, \textrm{GeV}) = 0.134$ 
as highlighted in blue in Tab.\ref{tab2-a2pi} and Tab.\ref{tab3-a2rho}. 

\begin{table}[tb]
\caption{The Gegenbauer moment $a_2^\rho$ at scale $\mu = 2 \, \textrm{GeV}$.}
\begin{center}
\begin{tabular}{c|c c c|c }
\toprule
\textrm{Method} & $f_{\rho}^T/f_{\rho}$  &  $a_2^{\rho,\parallel}(2 \, \textrm{GeV})$ & $a_2^{\rho,\perp}(2 \, \textrm{GeV})$ & \textrm{Refs} \\ 
\hline
QCDSR & 0.74(5)   & 0.11(5) & 0.11(5) & \cite{BallTB,BallNR} \\
\hline
LCSRs  & 0.751(7) & 0.17(7) & 0.14(6) & \cite{StraubICA} \\
\hline
LQCD        & 0.72(3) & - & - & \cite{BecirevicPN} \\
LQCD, QA & 0.742(14) & - & - &\cite{BraunJG} \\
LQCD, $N_f=2+1$, DWF & 0.687(27) & - & - & \cite{AlltonPN} \\
LQCD, $N_f=2+1$, QDF & - & 0.20(6) & - & \cite{ArthurXF} \\
{\blue LQCD, $N_f=2$} & {\blue 0.629(8)} & {\blue 0.132(27)} & {\blue 0.101(22)} & {\blue \cite{BraunWNX}} \\
\hline
\end{tabular}
\end{center}
\label{tab3-a2rho}
\end{table}

Now let us explain the derivation of the updated values in Tab.\ref{tab1-Bnl}. 
For the isovector 2$\pi$DAs, 
\begin{itemize}
\item[$\ddag$]
The soft pion theory, crossing relation and resonance approximation in Eqs.(\ref{eq:api-Bnl},\ref{eq:cr-vector},\ref{eq:arho-Bnl}) imply
\bea
&&B_{01}^{\parallel/\perp}(0) = a_0^{\pi/ \rho} = M_1^{\pi} = 1 \,, \,\,\,\,\,\, a_2^\pi = B_{21}^{\parallel}(0) + B_{23}^{\parallel}(0) \,, \non 
&&B_{21}^{\parallel/\perp}(0) = \frac{a_2^{\rho, \parallel/\perp}}{\text{Exp}[c_1^{\parallel/\perp,(21)} \, m_\rho^2]} \,,
\eea
to determine $B_{21}(0)$, we still acquiesce in the result for $c_1^{\parallel/\perp,(21)}$ obtained from the instanton model.
\item[$\ddag$]
Eq.\ref{eq:Bni-dr} and Eq.\ref{eq:frho-Bnl} are the main sources we used to determine the subtraction coefficients proportional 
to $k^2$, says $d/d k^2 \, \ln B_{01}(0)$.  
We take $g_{\rho\pi\pi} = 5.96$ from the energy-dependent $\rho \to \pi\pi$ width \cite{ChengSMJ}, 
use the experiment data of the scale-independent longitudinal decay constant $f_\rho = 0.21 \, \textrm{GeV}$, 
and adopt the relative ratio $f_\rho^T/f_\rho (\mu = 2 \, \textrm{GeV}) = 0.629$ evaluated from LQCD \cite{BraunWNX}. 
The running of the transversal decay constant of rho meson is the same as for the chirally odd dipion decay constant $f_{2\pi}^\perp$,
\be
f_\rho^T (\mu) = f_\rho^T (\mu_0) \left(\frac{\alpha_s(\mu)}{\alpha_s(\mu_0)} \right) ^{\gamma_0^{\perp,(0)}/(2\beta_0)}. 
\ee
\end{itemize}
While for the isoscalar 2$\pi$DAs, we have
\be
B_{10}^{\parallel}(0) + B_{12}^{\parallel}(0) = 0 \,, \,\,\,\,\,\,B_{12}^{\parallel}(0) = \frac{5}{9} M_2^\pi \sim \frac{5}{9} \,.
\ee
We find that the values of $d/d k^2 \, \ln B_{01}(0)$ determined by the decay constants of resonance 
closes to the result calculated from instanton model, 
which, conversely, supports our choice of $c_1^{(21)}$ from this model to predict $B_{nl}^{I=1}(0)$.  
The alteration of $B_{2l}(0)$ depends on the deep knowledge of $\pi$ and $\rho$ mesons ($a_2^\pi \,, a_2^\rho$).

\section{$B\to \pi\pi$ form factors}\label{sec:B2pipi_ff} 

We present the result of $B \to \pi\pi$ from factors with the updated subtraction constants in this section, 
in which the evolution of $F_{\parallel,\perp}^{I=1}$ on invariant mass and the form factors with isoscalar dipion state
$F^{I=0}$ are the new results, as the supplement to the previous work \cite{HambrockAOR,ChengSFK}. 

$B \to \pi\pi$ transition matrix element is defined in terms of the form factors as \cite{FallerDWA}, 
\bea
i\langle \pi^+(k_1) \pi^0(k_2)|\bar{u}\gamma_\nu(1-\gamma_5)b|\bar{B}^0(p)\rangle
= F_\perp (q^2, k^2,\zeta)\, \frac{2}{\sqrt{k^2} \sqrt{ \lambda_B}} \,
i\epsilon_{\nu\alpha\beta\gamma} \, q^\alpha \, k^{\beta} \, \bar{k}^{\gamma}
\nonumber \\
+ F_t (q^2,k^2,\zeta)\, \frac{q_\nu}{\sqrt{q^2}}
+ F_0 (q^2,k^2,\zeta)\, \frac{2\sqrt{q^2}}{\sqrt{\lambda_B}} \,
\Big(k_\nu - \frac{k \cdot q}{q^2} q_\nu\Big)
\nonumber\\
+ F_\parallel(q^2,k^2,\zeta) \, \frac{1}{\sqrt{k^2}} \,
\Big(\overline{k}_\nu - \frac{4 (q\cdot k) (q\cdot \overline{k})}{\lambda_B} \, k_\nu
+ \frac{4 k^2 (q\cdot \overline{k})}{\lambda_B} \, q_\nu \Big)\,,
\label{eq:ff-def}
\eea
where we use the same notations for the kinematics of dipion state as in Eqs.(\ref{eq:Phi-para},\ref{eq:Phi-perp}), 
$\overline{k}=k_1-k_2$.  
$\lambda_B = \lambda(m_B, q^2, k^2)$ is the K\"all\'en function ($\lambda(a,b,c) = a^2 +b^2 +c^2 -2ab-2ac-2bc$), 
$q=p-k$ indicates the momentum transfer in the decay, 
the vector products is expressed in terms of the independent variables as 
\be
2 q\cdot k = m_B^2 - q^2 -k^2 \,, \,\,\,\,\,\, 2 q\cdot \overline{k} = \sqrt{\lambda_B}\beta_\pi(k^2) \cos \theta_\pi = \sqrt{\lambda_B} (2\zeta-1) \,,
\ee 
with the phase factor $\beta_\pi(k^2) = \sqrt{1- 4m_\pi^2/k^2}$, and $\theta_\pi$ is the angle between the pions in their c.m. frame.
The derivation of $B \to \pi\pi$ from factors from LCSRs approach starts from defining an approximate correlation function, 
which is written down as the non-local matrix element with the $B$ meson interpolating current $j_5^{(B)}(0) = im_b\overline{b}(0)\gamma_5 q_{f'}(0)$ 
and the weak current $j_\mu^{V-A}(x)=\overline{q}_{f}(x)\gamma_\mu(1-\gamma_5)b(x)$, 
sandwiched between the vacuum and the on-shell dipion state,
\be
\Pi_\mu^{ab, ff'} (q,k_1,k_2) = i \int d^4x \, e^{iq\cdot x} \, \langle \pi^a(k_1) \pi^b(k_2) \vert T\left\{ j_\mu^{V-A}(x), j_5^{(B)}(0) \right\} \vert 0 \rangle \,.
\label{eq:cf-VA}
\ee 
This correlation function is only valid for calculating the form factors $F_{\perp}$ and $F_{\parallel}$,  
but is failed to derive the timelike-helicity ones ($F_{t}$ and $F_{0}$) due to the kinematic singularity, 
to overcome this problem we introduce an modified correlation with replacing the $V-A$ weak current by the 
pseudoscalar current $j^{(P)}(x) =  im_b\overline{q}_{f}(x)\gamma_5 b(x)$ \cite{ChengSFK}.

For the sake of simplicity, we focus on the form factors contributed at tree level\footnote{In the limit of massless lepton, 
timelike-helicity form factor $F_t^I$ does not contribute to the semileptonic $B \to \pi\pi \ell \overline{\nu}_{\ell}$ rate, 
but plays an important role in the factorization formula for nonleptonic three body $B$ decay, like $B \to \pi\pi\pi$.} 
in semileptonic decays $B \to \pi\pi \ell \overline{\nu}_\ell$, 
where the flavor of light antiquark in weak current is identified by f=u, 
and the quark in internal interpolating current is $f'=d$ and $u$ for $\{ab\} = \{+0\}$ and $\{00/+-\}$, respectively.
The QCD calculations and hadron analysis of the correlation functions are the same as in Refs. \cite{HambrockAOR,ChengSFK}, 
we here quote the result:
\bea
&&\frac{F_\parallel^{I}(q^2,k^2,\zeta)}{\sqrt{k^2}} = 
-\frac{m_b}{\sqrt{2}f_Bm_B^2 f_{2\pi}^\perp \, (2\zeta-1)} \int_{u_0}^1 \, \frac{du}{u^2} \, \Phi_{\perp}^{I}(u,\zeta,k^2) \, (m_b^2-q^2+k^2u^2)\, 
e^{-\frac{s(u)}{M^2}+\frac{m_B^2}{M^2}} \,, 
\label{eq:LCSR-Fpara} \\
&&\frac{F_\perp^{I}(q^2,k^2,\zeta)}{\sqrt{k^2} \sqrt{\lambda_B}} = 
\frac{m_b}{\sqrt{2}f_Bm_B^2 f_{2\pi}^\perp \, (2\zeta-1)} \int_{u_0}^1 \, \frac{du}{u} \, \Phi_{\perp}^{I}(u,\zeta,k^2) \, 
e^{-\frac{s(u)}{M^2}+\frac{m_B^2}{M^2}} \,, 
\label{eq:LCSR-Fperp} \\
&&\sqrt{q^2}F_t^I(q^2,k^2,\zeta) = - \frac{m_b^2}{\sqrt{2}f_Bm_B^2} \int_{u_0}^1 \, \frac{du}{u^2} \, \Phi_\parallel^I(u,\zeta,k^2) \,(m_b^2-q^2+k^2u^2)
e^{-\frac{s(u)}{M^2}+\frac{m_B^2}{M^2}} \, 
\label{eq:LCSR-Ft} \\
&&\sqrt{q^2} F_0^I(q^2,k^2,\zeta) = \frac{1}{m_B^2-q^2-k^2} \left[ \sqrt{\lambda_B}\sqrt{q^2} F_t(q^2,k^2,\zeta) + 
2 \sqrt{k^2} q^2 (2\zeta-1) F_\parallel^I(q^2,k^2,\zeta) \right] \,, 
\label{eq:LCSR-F0} 
\eea
where $s(u) = (m_B^2-q^2\overline{u}+k^2u\overline{u})/u$ and $u_0$ is the solution to $s(u_0) = s_0^B$, 
$M^2$ and $s_0^B$ are the Borel mass and threshold parameter introduced in LCSRs approach. 
Eqs.(\ref{eq:LCSR-Fperp}-\ref{eq:LCSR-F0}) collect the total contributions from all partial wave components, 
to obtain the contribution from each partial wave, we use the following expansion, 
\bea
&&F_{\parallel,\perp}(q^2,k^2,\zeta)
=\sum_{\ell=1}^\infty \,\sqrt{2\ell+1} \, F_{\parallel,\perp}^{(\ell)}(q^2,k^2) \,\frac{P_\ell^{(1)}(\cos\theta_\pi)}{\sin \theta_\pi} \,,
\label{eq:ffexpansionhel1} \\
&&F_{t,0}(q^2,k^2, \zeta)
=\sum_{\ell=0}^\infty \,\sqrt{2\ell+1} \, F_{t,0}^{(\ell)}(q^2,k^2) \, P_\ell^{(0)}(\cos\theta_\pi) \,.
\label{eq:ffexpansionhel2}
\eea

\subsection{Form factors with isovector dipion state}\label{subsec:ff_vector}

For the isovector dipion state, the partial wave contribution to the form factors is gained by multiplying both sides of 
Eqs.(\ref{eq:LCSR-Fpara},\ref{eq:LCSR-Fperp}) (Eq.\ref{eq:LCSR-Ft}) by $\sin \theta_\pi P_{\ell'}^{(1)}(\cos \theta_\pi)$ ($P_{\ell'}^{(0)}(\cos \theta_\pi)$)
and integrating over $\cos \theta_\pi$,
\bea
&&F_\parallel^{(\ell),I=1}(q^2,k^2) = \frac{\sqrt{k^2} \, m_b^3}{\sqrt{2} f_B m_B^2 \,  f_{2\pi}^\perp} \, e^{\frac{m_B^2}{M^2}} \, \sum_{\binom{n=0}{\textrm{even}}}^\infty \sum_{\binom{\ell'=1}{\textrm{odd}}}^{n+1} \, I_{\ell\ell'}^{I=1} \, B_{n\ell'}^{\perp, I=1} \, J_n^\parallel(q^2,k^2,M^2,s_0^B) \,, 
\label{eq:Fpara-vector} \\
&&F_\perp^{(\ell),I=1}(q^2,k^2) = \frac{\sqrt{k^2} \, \sqrt{\lambda_B} \, m_b}{\sqrt{2} f_B m_B^2 \, f_{2\pi}^\perp} \, e^{\frac{m_B^2}{M^2}} \, \sum_{\binom{n=0}{\textrm{even}}}^\infty \sum_{\binom{\ell'=1}{\textrm{odd}}}^{n+1} \, I_{\ell\ell'}^{I=1} \, B_{n\ell'}^{\perp, I=1} \, J_n^\perp(q^2,k^2,M^2,s_0^B) \,, 
\label{eq:Fperp-vector} \\
&&\sqrt{q^2} F_t^{(\ell),I=1}(q^2,k^2) = - \frac{m_b^4 }{\sqrt{2}f_Bm_B^2} \, \frac{\beta_\pi(k^2)} {\sqrt{2\ell+1}} \, e^{\frac{m_B^2}{M^2}}  \, 
\sum_{\binom{n=\ell-1}{\textrm{even}}}^{\infty} 
B_{n\ell}^{\parallel,I=1} \, J_n^t(q^2,k^2,M^2,s_0^B) \,,
\label{eq:Ft-vector}
\eea
where we introduce the short-hand notation for Legender integration, 
\bea
I_{\ell \ell'}^{I=1} \equiv -  \frac{\sqrt{2\ell+1} \, (\ell-1)!}{2(\ell+1)!} \int_{-1}^{1} \, \frac{d (\cos \theta_\pi)}{\cos \theta_\pi} \, \sqrt{1-\cos^2 \theta_\pi} 
P_{\ell}^{(1)}(\cos \theta_\pi) P_{\ell'}^{(0)}(\cos \theta_\pi) \,, 
\eea
and for the functions integrated over the momentum fraction, 
\bea
&&J_n^\parallel(q^2,k^2,M^2,s_0^B) \equiv  6 \int_{u_0}^1 \, \frac{du}{u} \, (1-u) \, C_n^{3/2}(2u-1) \,  (1 - \frac{q^2-u^2k^2}{m_b^2}) 
\, e^{-\frac{s(u)}{M^2}} \,,
\label{eq:Jpara} \\
&&J_n^\perp(q^2,k^2,M^2,s_0^B) \equiv 6 \int_{u_0}^1 \, du \, (1-u) \, C_n^{3/2}(2u-1) \, e^{-\frac{s(u)}{M^2}}\,,
\label{eq:Jperp} \\
&&J_n^t(q^2,k^2,M^2,s_0^B) \equiv J_n^\parallel(q^2,k^2,M^2,s_0^B) \,.
\label{eq:Jt}
\eea
To derive these expressions we use the orthogonality relation of the Legender polynomials 
\be
\int_{-1}^1\, P_\ell^n(x) \, P_{k}^n(x) = \frac{2}{2\ell+1}\, \frac{(\ell+n)!}{(\ell-n)!} \, \delta_{k\ell} \,.
\ee
In the pervious work\cite{HambrockAOR,ChengSFK}, 
these form factors are considered only around the threshold $k^2 \sim 4m_\pi^2$, 
we will replenish their evolution on $k^2$ to high energies in the next section with Eq.\ref{eq:Bni-dr} for the numerical computing.

\subsection{Form factors with isoscalar dipion state}\label{subsec:ff_scalar}

When the final two pions forms an isoscalar state, 
we multiply Eqs.(\ref{eq:LCSR-Fpara},\ref{eq:LCSR-Fperp}) by $\cos \theta_\pi P_{\ell'}^{(0)}(\cos \theta_\pi)$\footnote{
Multiplying by $\sin \theta_\pi P_{\ell'}^{(0)}(\cos \theta_\pi)$ will produce an imaginary part (kinematic singularity emerged 
from the integration over $\cos \theta_\pi$) due to the even number of $\ell'$ in isoscalar 2$\pi$DAs.} 
and Eq.\ref{eq:LCSR-Ft} by $P_{\ell'}^{(0)}(\cos \theta_\pi)$, 
then the $B^- \to \pi^0\pi^0$ form factors are arranged as
\bea
&&\sum_{\ell=1}^{\infty} \, I_{\ell \ell'}^{I=0} \, F_\parallel^{(\ell),I=0}(q^2,k^2) = \frac{\sqrt{k^2} \, m_b^3}{ f_B m_B^2 f_{2\pi}^\perp} \, 
e^{\frac{m_B^2}{M^2}} \, \sum_{\binom{n=1}{\textrm{odd}}}^{\infty} \sum_{\binom{\ell'=0}{\textrm{even}}}^{n+1} \frac{1}{2\ell'+1} 
B_{n \ell'}^{\perp,I=0}(k^2) \, J_n^{\parallel}(q^2,k^2,M^2,s_0^B) \,,
\label{eq:Fpara-scalar} \\
&&\sum_{\ell=1}^{\infty} \, I_{\ell \ell'}^{I=0} \, F_\perp^{(\ell),I=0}(q^2,k^2) = \frac{\sqrt{k^2} \, \sqrt{\lambda_B} \, m_b}{ f_B m_B^2 f_{2\pi}^\perp} \, 
e^{\frac{m_B^2}{M^2}}\, \sum_{\binom{n=1}{\textrm{odisod}}}^{\infty} \sum_{\binom{\ell'=0}{\textrm{even}}}^{n+1} \frac{1}{2\ell'+1} 
B_{n \ell'}^{\perp,I=0}(k^2) \, J_n^{\perp}(q^2,k^2,M^2,s_0^B) \,,
\label{eq:Fperp-scalar} \\
&&\sum_{\ell=0}^{\infty} \, \sqrt{2\ell+1} \, \sqrt{q^2} \, F_t^{(\ell),I=0}(q^2,k^2) = - \frac{m_b^4 \, \beta_\pi(k^2)}{2 f_B m_B^2} \, 
e^{\frac{m_B^2}{M^2}}\, \sum_{\binom{n=1}{\textrm{odd}}}^{\infty} \sum_{\binom{\ell=0}{\textrm{even}}}^{n+1} 
B_{n \ell}^{\parallel,I=0}(k^2) \, J_n^{t}(q^2,k^2,M^2,s_0^B) \,,
\label{eq:Ft-scalar} 
\eea 
the angular integration in this case is
\be
I_{\ell \ell'}^{I=0} = \sqrt{2\ell+1} \int_{-1}^1 \, d(\cos \theta_\pi) \, \frac{\cos \theta_\pi}{\sin \theta_\pi} \, 
P_{\ell'}^{(0)}(\cos \theta_\pi) \, P_{\ell}^{(1)}(\cos \theta_\pi) \,,
\ee
we reveal here that $I_{\ell \ell'}^{I=0} = 0$ when $\ell$ goes over odd numbers, 
$I_{20}^{I=0} = -2\sqrt{5} \,, I_{22}^{I=0} = -4/\sqrt{5}$ and $I_{2 \ell'}^{I=0} = 0$ when $\ell'>2$. 
Taking the accuracy at leading power for Eqs.(\ref{eq:Fpara-scalar},\ref{eq:Fperp-scalar}), 
which means neglecting the contributions from higher partial waves, 
we arrive at the final expression for $B^- \to \pi^0\pi^0$ form factors which had not been studied before, 
\bea
&&I_{2 \ell'}^{I=0} \, F_\parallel^{(\ell=2),I=0}(q^2,k^2) = \frac{\sqrt{k^2} \, m_b^3}{(2 \ell'+1) f_B m_B^2 f_{2\pi}^\perp} \, 
e^{\frac{m_B^2}{M^2}} \, \sum_{\binom{n=1}{\textrm{odd}}}^{\infty} \, B_{n \ell'}^{\perp,I=0}(k^2) \, J_n^{\parallel}(q^2,k^2,M^2,s_0^B) \,,
\label{eq:Fpara-scalar-D} \\
&&I_{2 \ell'}^{I=0} \, F_\perp^{(\ell=2),I=0}(q^2,k^2) = \frac{\sqrt{k^2} \, \sqrt{\lambda_B} \, m_b}{(2\ell'+1) f_B m_B^2 f_{2\pi}^\perp} \, 
e^{\frac{m_B^2}{M^2}}\, \sum_{\binom{n=1}{\textrm{odd}}}^{\infty} \, B_{n \ell'}^{\perp,I=0}(k^2) \, J_n^{\perp}(q^2,k^2,M^2,s_0^B) \,,
\label{eq:Fperp-scalar-D} \\
&&\sqrt{q^2} \, F_t^{(\ell=0),I=0}(q^2,k^2) = - \frac{m_b^4 \, \beta_\pi(k^2)}{2 f_B m_B^2} \, 
e^{\frac{m_B^2}{M^2}}\, \sum_{\binom{n=1}{\textrm{odd}}}^{\infty} \, B_{n 0}^{\parallel,I=0}(k^2) \, J_n^{t}(q^2,k^2,M^2,s_0^B) \,.
\label{eq:Ft-scalar-S} 
\eea
the comparison of Eq.\ref{eq:Fpara-scalar-D} and Eq.\ref{eq:Fperp-scalar-D} indicates the relation $2 B_{10}^{\perp}(k^2) = B_{12}^{\perp}(k^2)$ 
if we acquiesce in the convergence of gegenbauer expansion and keep only the first term with $n=1$, 
this relation leads to another constraint to check $\pi\pi$ phase shifts,
\be
\frac{d}{d k^2} \ln B_{10}^{\perp}(0) = \frac{2 k^2}{\pi} \, \int_{4m_\pi^2}^\infty \, ds \, \frac{\delta_2^0(s) - \delta_0^0(s)}{s^2 (s-k^2-i0)} \,.
\ee
As presented in Tab.\ref{tab1-Bnl}, the coeffficent $B_{10}^{\parallel}$ ($B_{12}^{\parallel}$) is also studied in instanton model, 
which allows us to predict the timelike-helicity form factor $\sqrt{q^2} \, F_t^{(\ell=0),I=0}(q^2,k^2)$, 
while the chirally odd coefficients are still missing for the other form factors in $B^- \to \pi^0\pi^0 \ell \overline{\nu}_\ell$ decay.


\section{Numerics}\label{sec:numerics}

\begin{figure}[tb]
\begin{center}
\vspace{-1cm}
\includegraphics[width=0.4\textwidth]{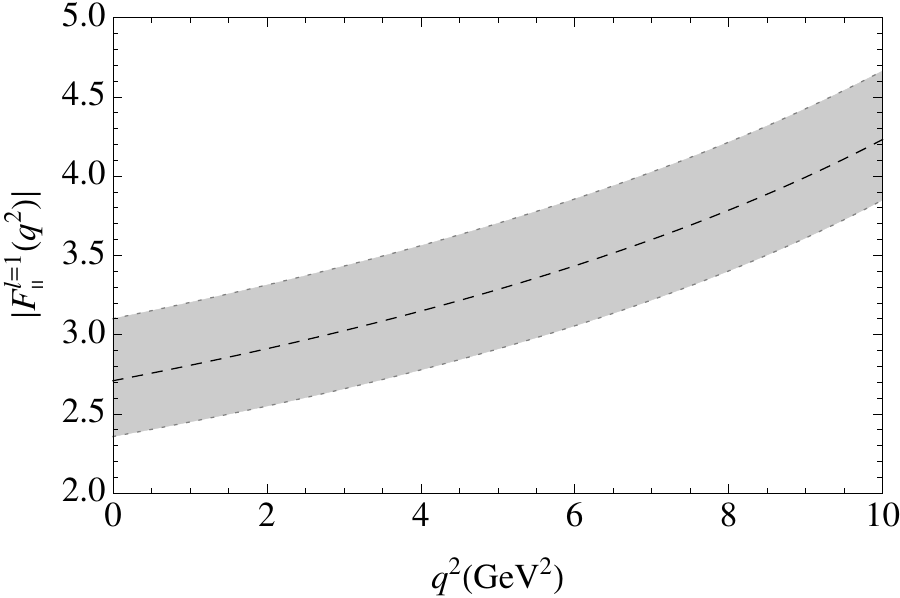}
\hspace{4mm}
\includegraphics[width=0.4\textwidth]{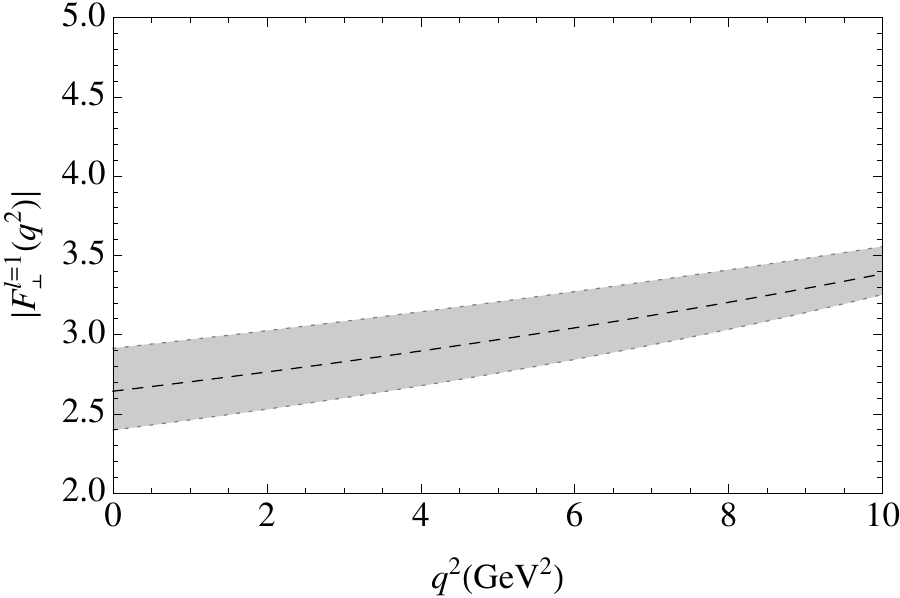}\non
\vspace{4mm}
\includegraphics[width=0.4\textwidth]{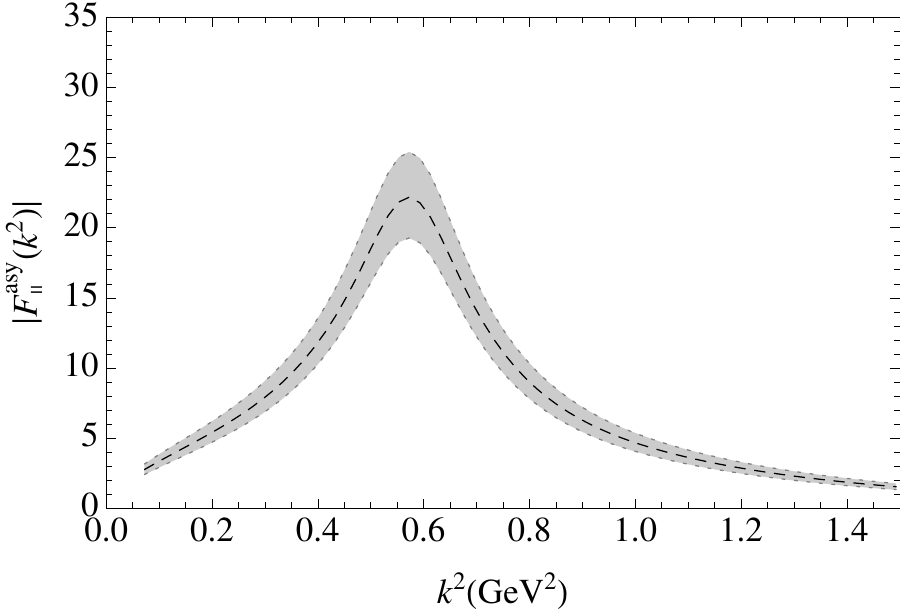}
\hspace{4mm}
\includegraphics[width=0.4\textwidth]{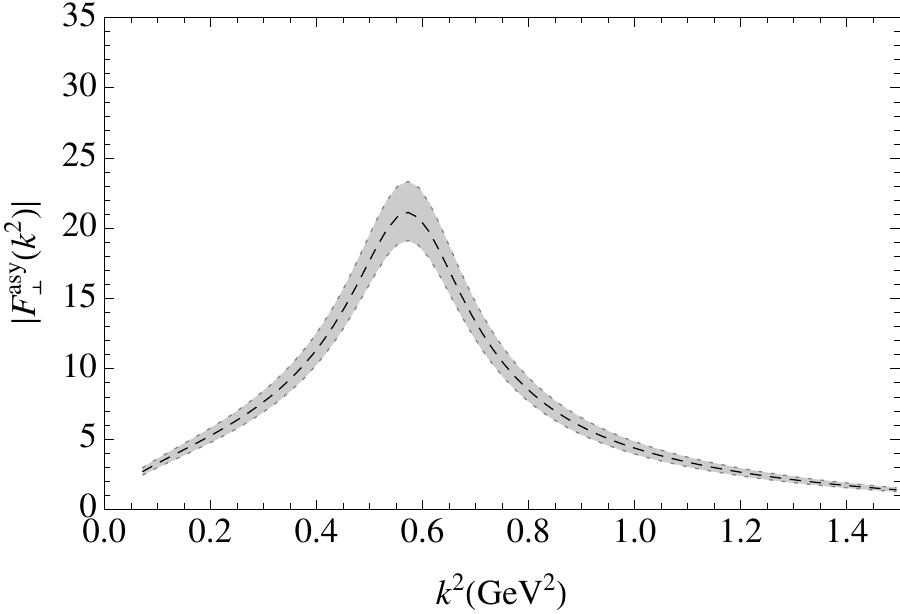}
\end{center}
\vspace{-0.6cm}
\caption{{\footnotesize $P-$wave contribution to $F_{\parallel,\perp}^{I=1}(q^2,k^2)$ 
at the scale $\mu = 3 \, \textrm{GeV}$ in Eqs.(\ref{eq:Fpara-vector},\ref{eq:Fperp-vector}).} }
\label{fig:1}
\end{figure}

To obtain the numerical result, we fix $b$ quark mass at $\overline{m}_b(3\, \textrm{GeV}) = 4.47 \, \textrm{GeV}$ \cite{TanabashiOCA}, 
and the decay constant of $B$ meson at $f_B = 0.207 \, \textrm{GeV}$ \cite{GelhausenWIA,AokiFRL}, with neglecting the small uncertainty from renormalization scale. 
For the LCSRs parameters we adopt the same inputs as in Refs.\cite{HambrockAOR,ChengSFK}: 
$M^2 = 16 \pm 4 \, \textrm{GeV}^2, \, s_0^B = 37.5 \pm 2.5 \, \textrm{GeV}^2$. 
Other parameters entered in the numerical computing are explained in Sec.\ref{sec:2piDAs}.

\begin{figure}[tb]
\begin{center}
\vspace{-1cm}
\includegraphics[width=0.4\textwidth]{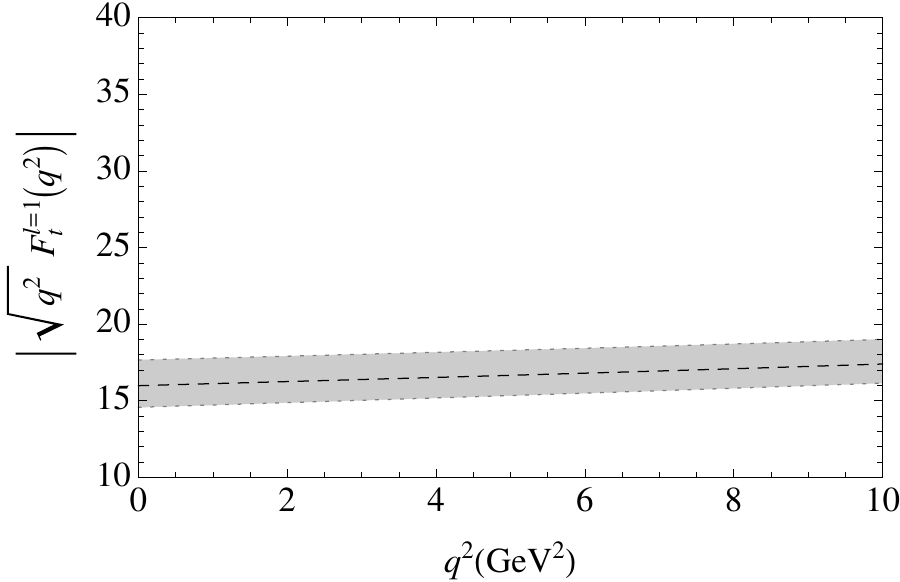}
\hspace{4mm}
\includegraphics[width=0.4\textwidth]{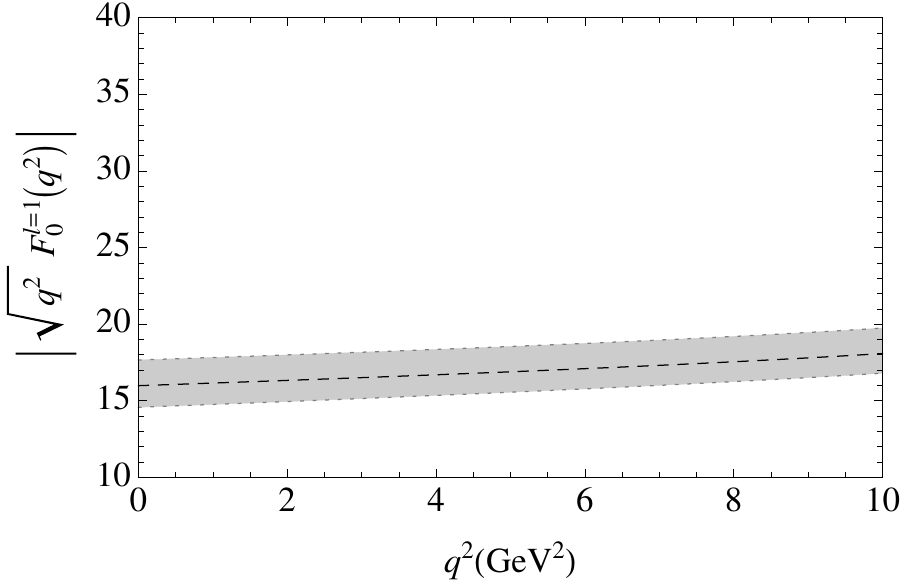}\non
\vspace{8mm}
\includegraphics[width=0.4\textwidth]{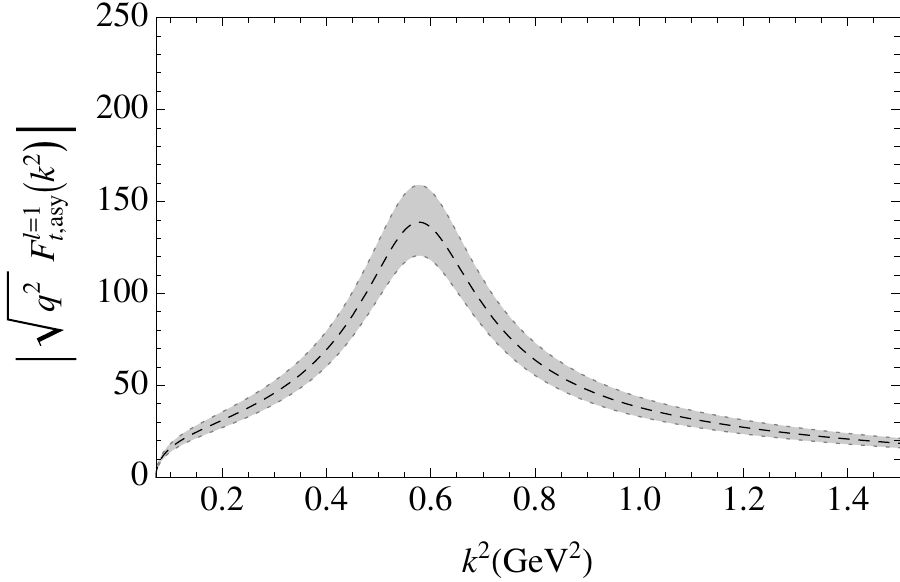}
\end{center}
\vspace{-0.6cm}
\caption{{\footnotesize $P-$wave contribution to $\sqrt{q^2}F_{t,0}^{I=1}(q^2,k^2)$ 
at the scale $\mu = 3 \, \textrm{GeV}$ in Eqs.(\ref{eq:Ft-vector},\ref{eq:LCSR-F0}).} }
\label{fig:2}
\end{figure}

\begin{figure}[tb]
\begin{center}
\vspace{0.5cm}
\includegraphics[width=0.4\textwidth]{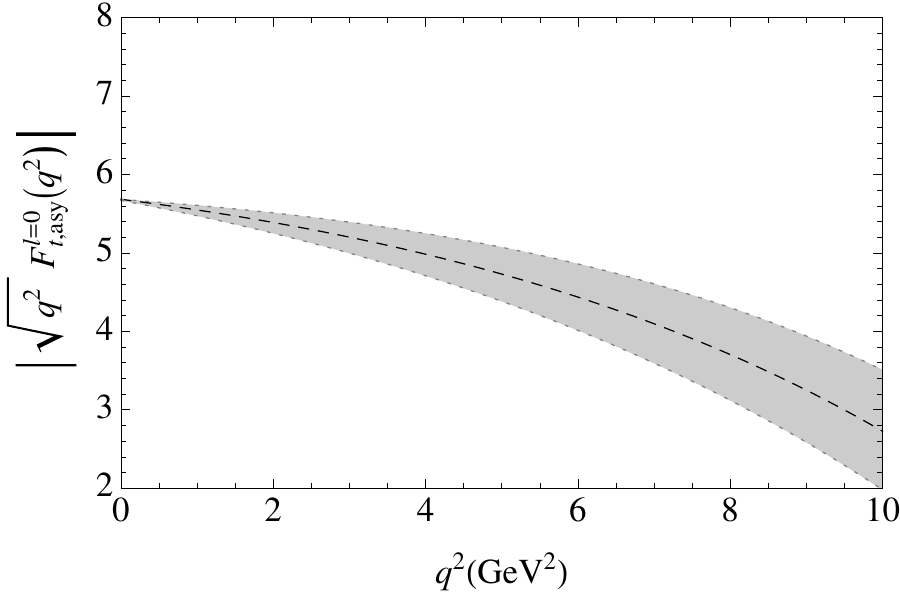}
\vspace{4mm}
\includegraphics[width=0.4\textwidth]{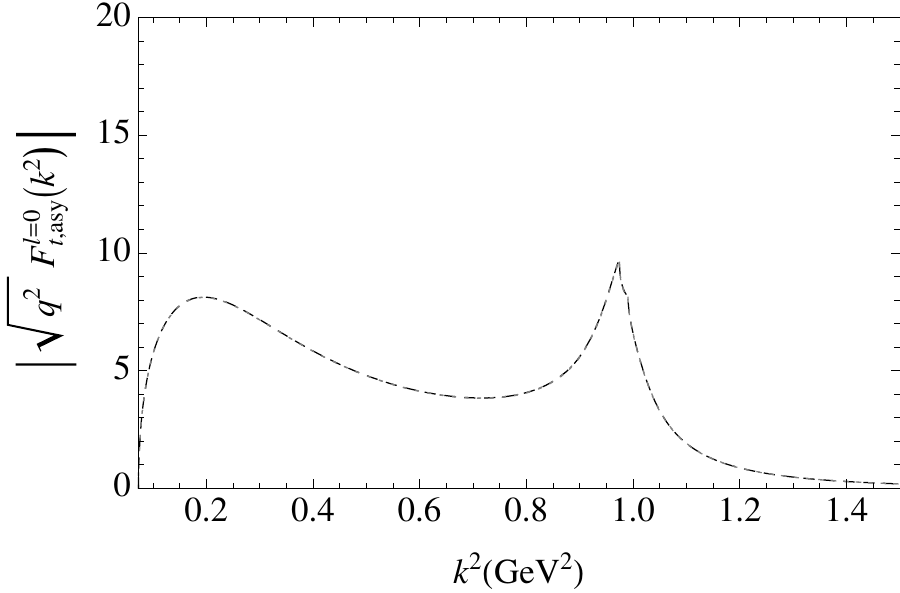}
\end{center}
\vspace{-1cm}
\caption{{\footnotesize $S-$wave contribution to $\sqrt{q^2}F_{t}^{I=0}(q^2,k^2)$ 
at the scale $\mu = 3 \, \textrm{GeV}$ in Eq.(\ref{eq:Ft-scalar-S}).} }
\label{fig:3}
\end{figure}

In Figs.(\ref{fig:1}-\ref{fig:3}) we present the LCSRs prediction for $B \to \pi \pi$ form factors, 
in which we adopt the notations $F_{\parallel,\perp}(q^2) = F_{\parallel,\perp}(q^2,k^2 = 4m_\pi^2), \, 
F_{t,0}(q^2) = F_{t,0}(q^2,k^2 = 0.1 \, \textrm{GeV}^2)$ and $F(k^2) = F(q^2=0,k^2)$. 
For the transition form factors with isovector dipion state, 
the $P-$wave contributions to $F(q^2)$ are calculated at low $k^2$ with the few first expansion coefficients of
$B_{nl}(k^2) = B_{nl}(0) + k^2 d/dk^2 \ln B_{nl}(0)$ listed in Tab.\ref{tab1-Bnl}, 
while the curves of $F_{asy}(k^2)$ are acquired by using the Omn\'es solution of $B_{01}(k^2)$ expressed in Eq.\ref{eq:Bni-dr}.  
When the final dipion is isoscalar state, we plot only the asymptotic shapes of the form factors because
we do not have any information so far beyond the asymptotic coefficient $B_{10}(k^2)$ from experiment measurement or from effective low energy theory. 
The resonance information in the invariant mass spectrum is carried by the pion-pion phase shift $\delta_\ell^I(k^2)$, 
which is well measured and described by the Regge parameterization in the range from the threshold to $1100 \, \textrm{MeV}$ \cite{GarciaMartinCN}.  
This is not enough because the coefficients $B_{nl}(k^2)$ in Eq.\ref{eq:Bni-dr} integrate over the whole region of invariant mass suqared,  
we adopt the result from amplitude analysis with marriage of dispersion relations with unitarity\cite{DaiZTA,DaiUAO}, 
which it is able to extrapolate the phase shifts to a high energy $\sim 5 \, \textrm{GeV}$ for both $S-$wave and $P-$wave, 
with considering all the well measured data, the $\pi\pi-KK$ final state interaction, 
the mass difference between charged and neutral Kaon and also the low energy Roy-Equation.

The upper panels of Figs.(\ref{fig:1},\ref{fig:2}) show the $P-$wave contribution to $q^2-$dependence form factors of 
$\overline{B}^0 \to \pi^+ \pi^0 \ell \overline{\nu}_\ell$ decay around the dipion threshold, 
consisting with our previous work \cite{HambrockAOR,ChengSFK} with in the uncertainty analysis.  
The lower panels are the new result for the form factors in a wide range of invariant mass squared at the full recoil point, 
the shape of timelike-helicity form factor $F_t(k^2)$ is compatible with it obtained\cite{ChengSFK} by using 
the normalization condition $B_{01}^\parallel(k^2) = F_\pi(k^2)$
and the data of pion form factor measured up to $1.78 \, \textrm{GeV}$ \cite{FujikawaMA}, 
which in turn supports that Eq.\ref{eq:Bni-dr} is powerful at least in the few low resonances interval. 
Comparing with the result obtained from LCSRs with the $B-$meson distribution amplitudes\cite{ChengSMJ}, 
the $\sim 25 \%$ differences is regarded as the contributions from high twist 2$\pi$DAs. 
Fig.\ref{fig:3} depicts the first calculation of $B^- \to \pi^0\pi^0$ form factors, 
where the $S-$wave contribution is shown.
It is obvious to see that the $q^2-$dependence of $F_t^{I=0}$ is decreasing, in contrasting to $F_t^{I=1}$, 
which is originated from the different gegenbauer polynomials $C_n^{3/2}(2 u -1)$ in function $J_n^t$. 
In the right panel for the $k^2-$dependence, the uncertainties from LCSRs parameters 
cannel between the exponential $e^{m_B^2/M^2}$ and the function $J_0^t(M^2, s_0^B)$. 

\section{Conclusion and outlook}\label{sec:conclusion}

In this paper, we discuss and update the isovector 2$\pi$DAs  by their relations to the distribution amplitudes of pion and rho mesons, 
with which we revisit the $\overline{B}^0 \to \pi^+\pi^0$ form factors from LCSRs approach. 
With the Omn\'es solution of expansion coefficients of 2$\pi$DAs in terms of the $\pi\pi$ phase shifts and a few subtraction constants, 
the LCSRs predictions for these form factors are extended from the threshold of dipion invariant mass  
to a wide range energies with including resonances.  
We also study the $B^- \to \pi^0\pi^0$ form factors with the isoscalar 2$\pi$DAs 
calculated from effective low energy theory based on instanton vacuum. 
For the form factors with isovector dipion state, the updated calculation does not bring noticeable deviation from the previous work, 
the lowest intermediate resonance $\rho$ dominates in $P-$wave contribution ($\sim 80 \%$), 
the high partial ($l=3$) contribution is tiny and not exceed a few percents of the $P-$wave contribution, 
and the contribution from high gegenbauer term ($n=2$) is barely too. 
For the form factors with isoscalar dipion state, our calculation is asymptotic 
with neglecting the contributions from higher partial waves and lacking of the information for high gegenbauer terms, 
the contribution from resonance in $S-$wave is not apparent as in the $P-$wave, 
indicating a more complicated inner structure in isoscalar dipion system. 

Further improvements on this project include:
(a) Developing and promoting the effective low energy theory and/or other approaches 
to calculate the chirally odd coefficients of isoscalar 2$\pi$DAs, 
also for the higher power terms of chirally even coefficients.
(b) Finding the relation between isoscalar dipion state and intermediate meson state ($f_0$) 
to restrict the subtraction constants in isoscalar 2$\pi$DAs.
(c) Forwarding the calculation to include the contributions from 2$\pi$DAs at twist-3 and from the next-to-leading-order QCD correction, 
to meet the precision requirement for extracting the CKM matrix element $V_{ub}$. 
(d) Revisiting these form factors from another LCSRs with $B$ meson DAs as input, 
in order to improve the theoretical accuracy.

\section*{Acknowledgments}
We are grateful to V. M. Braun, M.V. Polyakov for valuable discussions, 
and to A. Khodjamirian and J. Virto for the previous collaborations, 
we also wish to thank Ling-yun Dai for the discussion on the pion-pion phase shifts and provide the original result from amplitude analysis. 
This work is supported by the National Science Foundation of China under the No.11805060 
and "the Fundamental Research Funds for the Central Universities" under No 020400/531107051171. 




\begin{thebibliography}{99}

\bibitem{FallerDWA}
  S.~Faller, T.~Feldmann, A.~Khodjamirian, T.~Mannel and D.~van Dyk,
  \textit{Disentangling the Decay Observables in $B^- \to \pi^+\pi^-\ell^-\bar\nu_\ell$},
Phys.\ Rev.\ D {\bf 89}, no. 1, 014015 (2014), 
[arXiv:1310.6660 [hep-ph]].

\bibitem{BoerIEZ}
  P.~Böer, T.~Feldmann and D.~van Dyk,
   \textit{QCD Factorization Theorem for $B \to \pi\pi\ell\nu$ Decays at Large Dipion Masses},
JHEP {\bf 1702}, 133 (2017), 
[arXiv:1608.07127 [hep-ph]].

\bibitem{FeldmannKQR}
  T.~Feldmann, D.~Van Dyk and K.~K.~Vos,
   \textit{Revisiting $B \to \pi\pi \ell \nu$ at large dipion masses},
JHEP {\bf 1810}, 030 (2018), 
[arXiv:1807.01924 [hep-ph]].

\bibitem{KangJAA}
  X.~W.~Kang, B.~Kubis, C.~Hanhart and U.~G.~Meißner,
   \textit{$B_{l4}$ decays and the extraction of $|V_{ub}|$},
Phys.\ Rev.\ D {\bf 89}, 053015 (2014), 
[arXiv:1312.1193 [hep-ph]].

\bibitem{MeissnerHYA}
  U.~G.~Mei\ss ner and W.~Wang,
     \textit{Generalized Heavy-to-Light Form Factors in Light-Cone Sum Rules},
Phys.\ Lett.\ B {\bf 730}, 336 (2014), 
[arXiv:1312.3087 [hep-ph]].

\bibitem{MeissnerPBA}
  U.~G.~Mei\ss ner and W.~Wang,
     \textit{${\bf B_s\to K^{(*)} \ell\bar \nu}$, Angular Analysis, S-wave Contributions and ${\bf |V_{ub}|}$}, 
JHEP {\bf 1401}, 107 (2014), 
[arXiv:1311.5420 [hep-ph]].

\bibitem{HambrockAOR}
  C.~Hambrock and A.~Khodjamirian,
   \textit{Form factors in $\bar B^0 \to \pi^+\pi^0 \ell \bar\nu_\ell$ from QCD light-cone sum rules},
Nucl.\ Phys.\ B {\bf 905}, 373 (2016), 
[arXiv:1511.02509 [hep-ph]].

\bibitem{ChengSFK}
  S.~Cheng, A.~Khodjamirian and J.~Virto,
   \textit{Timelike-helicity $B\to \pi\pi$ form factor from light-cone sum rules with dipion distribution amplitudes},
Phys.\ Rev.\ D {\bf 96}, no. 5, 051901 (2017),
[arXiv:1709.00173 [hep-ph]].

\bibitem{ChengSMJ}
  S.~Cheng, A.~Khodjamirian and J.~Virto,
   \textit{$B\to\pi\pi$ Form Factors from Light-Cone Sum Rules with $B$-meson Distribution Amplitudes},
JHEP {\bf 1705}, 157 (2017),
[arXiv:1701.01633 [hep-ph]].

\bibitem{PolyakovZE}
  M.~V.~Polyakov,
   \textit{Hard exclusive electroproduction of two pions and their resonances},
Nucl.\ Phys.\ B {\bf 555}, 231 (1999), 
[hep-ph/9809483].

\bibitem{BaierVW}
  V.~N.~Baier and A.~G.~Grozin,
   \textit{Hadron Cluster Production in Hard Process},
   Sov.\ J.\ Nucl.\ Phys.\  {\bf 35}, 899 (1982), [Yad.\ Fiz.\  {\bf 35}, 1537 (1982)].

\bibitem{Grozin:1983tt}
   A.~G.~Grozin,
   \textit{On Wave Functions Of Mesonic Pairs And Mesonic Resonances},
   Sov.\ J.\ Nucl.\ Phys.\  {\bf 38} (1983) 289
    [Yad.\ Fiz.\  {\bf 38} (1983) 484].

\bibitem{Grozin:1986at}
   A.~G.~Grozin,
      \textit{One And Two Particle Wave Functions Of Multi-Hadron Systems},
   Theor.\ Math.\ Phys.\  {\bf 69} (1986) 1109
    [Teor.\ Mat.\ Fiz.\  {\bf 69} (1986) 219].
   
\bibitem{Diehl-1998}
  M.~Diehl, T.~Gousset, B.~Pire and O.~Teryaev,
   \textit{Probing partonic structure in $\gamma^\ast \gamma \to \pi \pi$ near threshold},
  Phys.\ Rev.\ Lett.\  {\bf 81}, 1782 (1998), 
  [hep-ph/9805380].
  
\bibitem{Diehl-2000}
  M.~Diehl, T.~Gousset and B.~Pire,
   \textit{Exclusive production of pion pairs in $\gamma^\ast \gamma$ collisions at large $Q^2$},
  Phys.\ Rev.\ D {\bf 62}, 073014 (2000), 
  [hep-ph/0003233].

\bibitem{Diehl-2002}
 P.~Hägler, B.~Pire, L.~Szymanowski and O.~V.~Teryaev,
   \textit{Hunting the QCD-Odderon in hard diffractive electroproduction of two pions},
  Phys.\ Lett.\ B {\bf 535}, 117 (2002), Erratum: [Phys.\ Lett.\ B {\bf 540}, 324 (2002)], 
  [hep-ph/0202231].
   
\bibitem{Diehl-2003}
B.~Pire and L.~Szymanowski,
  \textit{Impact representation of generalized distribution amplitudes}, 
  Phys.\ Lett.\ B {\bf 556}, 129 (2003), 
  [hep-ph/0212296].

\bibitem{LehmannDronkeAQ}
  B.~Lehmann-Dronke, P.~V.~Pobylitsa, M.~V.~Polyakov, A.~Schafer and K.~Goeke,
   \textit{Hard diffractive electroproduction of two pions},
Phys.\ Lett.\ B {\bf 475}, 147 (2000), 
[hep-ph/9910310].

\bibitem{JiPC}
  X.~D.~Ji,
   \textit{Off forward parton distributions},
J.\ Phys.\ G {\bf 24}, 1181 (1998), 
[hep-ph/9807358].

\bibitem{PolyakovEXB}
  M.~V.~Polyakov and H.~D.~Son,
   \textit{Nucleon gravitational form factors from instantons: forces between quark and gluon subsystems},
JHEP {\bf 1809}, 156 (2018), 
[arXiv:1808.00155 [hep-ph]].

\bibitem{ChernyakAS}
  V.~L.~Chernyak and A.~R.~Zhitnitsky,
   \textit{Asymptotic Behavior of Hadron Form-Factors in Quark Model. (In Russian)},
JETP Lett.\  {\bf 25}, 510 (1977), [Pisma Zh.\ Eksp.\ Teor.\ Fiz.\  {\bf 25}, 544 (1977)].

\bibitem{LepageZB}
  G.~P.~Lepage and S.~J.~Brodsky,
   \textit{Exclusive Processes in Quantum Chromodynamics: Evolution Equations for Hadronic Wave Functions and the Form-Factors of Mesons},
Phys.\ Lett.\  {\bf 87B}, 359 (1979).

\bibitem{GrossCS}
  D.~J.~Gross and F.~Wilczek,
   \textit{Asymptotically Free Gauge Theories. 2.},
Phys.\ Rev.\ D {\bf 9}, 980 (1974).

\bibitem{PolyakovTD}
  M.~V.~Polyakov and C.~Weiss,
   \textit{Two pion light cone distribution amplitudes from the instanton vacuum},
Phys.\ Rev.\ D {\bf 59}, 091502 (1999), 
[hep-ph/9806390].


\bibitem{ChernyakEJ}
  V.~L.~Chernyak and A.~R.~Zhitnitsky,
   \textit{Asymptotic Behavior of Exclusive Processes in QCD},
Phys.\ Rept.\  {\bf 112}, 173 (1984).

\bibitem{ChernyakZZ}
  V.~L.~Chernyak and A.~R.~Zhitnitsky,
   \textit{Exclusive Decays of Heavy Mesons},
Nucl.\ Phys.\ B {\bf 201}, 492 (1982), Erratum: [Nucl.\ Phys.\ B {\bf 214}, 547 (1983)].

\bibitem{KhodjamirianGA}
  A.~Khodjamirian, T.~Mannel and M.~Melcher,
   \textit{Kaon distribution amplitude from QCD sum rules},
Phys.\ Rev.\ D {\bf 70}, 094002 (2004), 
[hep-ph/0407226].

\bibitem{BallWN}
  P.~Ball, V.~M.~Braun and A.~Lenz,
  \textit{ Higher-twist distribution amplitudes of the K meson in QCD},
JHEP {\bf 0605}, 004 (2006), 
[hep-ph/0603063].

\bibitem{MikhailovPT}
  S.~V.~Mikhailov and A.~V.~Radyushkin,
   \textit{The Pion wave function and QCD sum rules with nonlocal condensates},
Phys.\ Rev.\ D {\bf 45}, 1754 (1992).

\bibitem{BakulevPF}
  A.~P.~Bakulev and S.~V.~Mikhailov,
   \textit{The rho meson and related meson wave functions in QCD sum rules with nonlocal condensates},
Phys.\ Lett.\ B {\bf 436}, 351 (1998), 
[hep-ph/9803298].

\bibitem{SchmeddingAP}
  A.~Schmedding and O.~I.~Yakovlev,
   \textit{Perturbative effects in the form-factor $\gamma \gamma^{\ast} \to \pi^0$ and extraction of the pion wave function from CLEO data},
Phys.\ Rev.\ D {\bf 62}, 116002 (2000), 
[hep-ph/9905392].

\bibitem{BakulevUC}
  A.~P.~Bakulev, S.~V.~Mikhailov and N.~G.~Stefanis,
   \textit{Unbiased analysis of CLEO data at NLO and pion distribution amplitude},
Phys.\ Rev.\ D {\bf 67}, 074012 (2003), 
[hep-ph/0212250].

\bibitem{AgaevRC}
  S.~S.~Agaev,
   \textit{Impact of the higher twist effects on the $\gamma \gamma^{\ast} \to \pi^0$ transition form-factor},
Phys.\ Rev.\ D {\bf 72}, 114010 (2005), Erratum: [Phys.\ Rev.\ D {\bf 73}, 059902 (2006)], 
[hep-ph/0511192].

\bibitem{BakulevCP}
  A.~P.~Bakulev, S.~V.~Mikhailov and N.~G.~Stefanis,
   \textit{Tagging the pion quark structure in QCD},
Phys.\ Rev.\ D {\bf 73}, 056002 (2006), 
[hep-ph/0512119].

\bibitem{AgaevAQ}
  S.~S.~Agaev, V.~M.~Braun, N.~Offen and F.~A.~Porkert,
   \textit{Light Cone Sum Rules for the $\pi^0\gamma^{\ast}\gamma$ Form Factor Revisited},
Phys.\ Rev.\ D {\bf 83}, 054020 (2011), 
[arXiv:1012.4671 [hep-ph]].

\bibitem{AgaevTM}
  S.~S.~Agaev, V.~M.~Braun, N.~Offen and F.~A.~Porkert,
   \textit{BELLE Data on the $\pi^0 \gamma^\ast \gamma$ Form Factor: A Game Changer?},
Phys.\ Rev.\ D {\bf 86}, 077504 (2012), 
[arXiv:1206.3968 [hep-ph]].

\bibitem{BraunUJ}
  V.~M.~Braun, A.~Khodjamirian and M.~Maul,
   \textit{Pion form-factor in QCD at intermediate momentum transfers},
Phys.\ Rev.\ D {\bf 61}, 073004 (2000), 
[hep-ph/9907495].

\bibitem{AgaevGU}
  S.~S.~Agaev,
   \textit{Higher twist distribution amplitudes of the pion and electromagnetic form-factor $F_{\pi} (Q^2)$},
Phys.\ Rev.\ D {\bf 72}, 074020 (2005), 
[hep-ph/0509345].

\bibitem{BallTB1}
  P.~Ball and R.~Zwicky,
   \textit{$\vert V_{ub} \vert $ and constraints on the leading-twist pion distribution amplitude from $B \to \pi l \nu$},
Phys.\ Lett.\ B {\bf 625}, 225 (2005), 
[hep-ph/0507076].

\bibitem{DuplancicIX}
  G.~Duplancic, A.~Khodjamirian, T.~Mannel, B.~Melic and N.~Offen,
   \textit{Light-cone sum rules for $B \to \pi$ form factors revisited},
JHEP {\bf 0804}, 014 (2008), 
[arXiv:0801.1796 [hep-ph]].

\bibitem{KhodjamirianUB}
  A.~Khodjamirian, T.~Mannel, N.~Offen and Y.-M.~Wang,
   \textit{$B \to \pi \ell \nu_l$ Width and $|V_{ub}|$ from QCD Light-Cone Sum Rules},
Phys.\ Rev.\ D {\bf 83}, 094031 (2011), 
[arXiv:1103.2655 [hep-ph]].

\bibitem{BraunDG}
  V.~M.~Braun {\it et al.},
   \textit{Moments of pseudoscalar meson distribution amplitudes from the lattice},
Phys.\ Rev.\ D {\bf 74}, 074501 (2006), 
[hep-lat/0606012].

\bibitem{ArthurXF}
  R.~Arthur, P.~A.~Boyle, D.~Brommel, M.~A.~Donnellan, J.~M.~Flynn, A.~Juttner, T.~D.~Rae and C.~T.~C.~Sachrajda,
   \textit{Lattice Results for Low Moments of Light Meson Distribution Amplitudes},
Phys.\ Rev.\ D {\bf 83}, 074505 (2011), 
[arXiv:1011.5906 [hep-lat]].

\bibitem{BraunAXA}
  V.~M.~Braun, S.~Collins, M.~Göckeler, P.~Pérez-Rubio, A.~Schäfer, R.~W.~Schiel and A.~Sternbeck,
   \textit{Second Moment of the Pion Light-cone Distribution Amplitude from Lattice QCD},
Phys.\ Rev.\ D {\bf 92}, no. 1, 014504 (2015), 
[arXiv:1503.03656 [hep-lat]].

\bibitem{BaliUDE}
  G.~S.~Bali {\it et al.} [RQCD Collaboration],
   \textit{Second moment of the pion distribution amplitude with the momentum smearing technique},
Phys.\ Lett.\ B {\bf 774}, 91 (2017), 
[arXiv:1705.10236 [hep-lat]].

\bibitem{BraunWNX}
  V.~M.~Braun {\it et al.},
   \textit{The $\rho$-meson light-cone distribution amplitudes from lattice QCD},
JHEP {\bf 1704}, 082 (2017), 
[arXiv:1612.02955 [hep-lat]].


\bibitem{BallTB}
  P.~Ball and V.~M.~Braun,
   \textit{The Rho meson light cone distribution amplitudes of leading twist revisited},
Phys.\ Rev.\ D {\bf 54}, 2182 (1996), 
[hep-ph/9602323].

\bibitem{BallNR}
  P.~Ball and R.~Zwizky,
   \textit{$|V_{td} / V_{ts}|$ from $B \to V \gamma$},
JHEP {\bf 0604}, 046 (2006), 
[hep-ph/0603232].

\bibitem{StraubICA}
  A.~Bharucha, D.~M.~Straub and R.~Zwicky,
   \textit{$B\to V\ell^+\ell^-$ in the Standard Model from light-cone sum rules},
JHEP {\bf 1608}, 098 (2016), 
[arXiv:1503.05534 [hep-ph]].

\bibitem{BecirevicPN}
  D.~Becirevic, V.~Lubicz, F.~Mescia and C.~Tarantino,
   \textit{Coupling of the light vector meson to the vector and to the tensor current},
JHEP {\bf 0305}, 007 (2003), 
[hep-lat/0301020].

\bibitem{BraunJG}
  V.~M.~Braun, T.~Burch, C.~Gattringer, M.~Gockeler, G.~Lacagnina, S.~Schaefer and A.~Schafer,
   \textit{A Lattice calculation of vector meson couplings to the vector and tensor currents using chirally improved fermions},
Phys.\ Rev.\ D {\bf 68}, 054501 (2003), 
[hep-lat/0306006].

\bibitem{AlltonPN}
  C.~Allton {\it et al.} [RBC-UKQCD Collaboration],
   \textit{Physical Results from 2+1 Flavor Domain Wall QCD and SU(2) Chiral Perturbation Theory},
Phys.\ Rev.\ D {\bf 78}, 114509 (2008), 
[arXiv:0804.0473 [hep-lat]].

\bibitem{TanabashiOCA}
  M.~Tanabashi {\it et al.} [Particle Data Group],
   \textit{Review of Particle Physics},
Phys.\ Rev.\ D {\bf 98}, no. 3, 030001 (2018).

\bibitem{GelhausenWIA}
  P.~Gelhausen, A.~Khodjamirian, A.~A.~Pivovarov and D.~Rosenthal,
   \textit{Decay constants of heavy-light vector mesons from QCD sum rules},
Phys.\ Rev.\ D {\bf 88}, 014015 (2013), Erratum: [Phys.\ Rev.\ D {\bf 89}, 099901 (2014)], Erratum: [Phys.\ Rev.\ D {\bf 91}, 099901 (2015)], 
[arXiv:1305.5432 [hep-ph]].

\bibitem{AokiFRL}
  S.~Aoki {\it et al.},
   \textit{Review of lattice results concerning low-energy particle physics},
Eur.\ Phys.\ J.\ C {\bf 77}, no. 2, 112 (2017), 
[arXiv:1607.00299 [hep-lat]].

\bibitem{GarciaMartinCN}
  R.~Garcia-Martin, R.~Kaminski, J.~R.~Pelaez, J.~Ruiz de Elvira and F.~J.~Yndurain,
   \textit{The Pion-pion scattering amplitude. IV: Improved analysis with once subtracted Roy-like equations up to 1100 MeV},
Phys.\ Rev.\ D {\bf 83}, 074004 (2011), 
[arXiv:1102.2183 [hep-ph]].

\bibitem{DaiZTA}
  L.~Y.~Dai and M.~R.~Pennington,
   \textit{Comprehensive amplitude analysis of $\gamma\gamma \rightarrow \pi^+\pi^-, \pi^0\pi^0$ and $\overline{K} K$ below 1.5 GeV},
Phys.\ Rev.\ D {\bf 90}, no. 3, 036004 (2014), 
[arXiv:1404.7524 [hep-ph]].

\bibitem{DaiUAO}
  L.~Y.~Dai and U.~G.~Meißner,
   \textit{A new method to study the number of colors in the final-state interactions of hadrons},
Phys.\ Lett.\ B {\bf 783}, 294 (2018), 
[arXiv:1706.10123 [hep-ph]].

\bibitem{FujikawaMA}
  M.~Fujikawa {\it et al.} [Belle Collaboration],
   \textit{High-Statistics Study of the $\tau \to \pi^- \pi^0 \nu_{\tau}$ Decay},
Phys.\ Rev.\ D {\bf 78}, 072006 (2008), 
[arXiv:0805.3773 [hep-ex]].









\end{thebibliography}
\end{document}